\documentclass[twocolumn]{aastex61}
\pdfoutput=1 
\usepackage{amsmath,amstext}
\usepackage[T1]{fontenc}
\usepackage{apjfonts} 
\usepackage[figure,figure*]{hypcap}
\usepackage{lineno}

\newcommand{\gw}{GW170817}
\newcommand{\rp}{$r$-process}
\newcommand{\kms}{km~s$^{-1}$} 
\newcommand{\msun}{M$_{\odot}$}
\newcommand{\xlan}{$X_{\mathrm{lan}}$}

\shorttitle{NIR Spectroscopy of EM Counterpart to \gw}
\shortauthors{Chornock et al.}

\begin{document}

\title{The Electromagnetic Counterpart of the Binary Neutron Star Merger LIGO/VIRGO \gw. IV. Detection of Near-Infrared Signatures of {\it \MakeLowercase{r}}-process Nucleosynthesis With Gemini-South}

\author{R. Chornock}
\affiliation{Astrophysical Institute, Department of Physics and Astronomy, 251B Clippinger Lab, Ohio University, Athens, OH 45701, USA; \texttt{chornock@ohio.edu}}

\author{E. Berger}
\affiliation{Harvard-Smithsonian Center for
Astrophysics, 60 Garden Street, Cambridge, MA 02138, USA}
\author{D. Kasen}
\affil{Departments of Physics and Astronomy, and Theoretical Astrophysics Center, University of California, Berkeley, CA 94720-7300, USA}
\affiliation{Nuclear Science Division, Lawrence Berkeley National Laboratory, Berkeley, CA 94720-8169, USA
}
\author{P.~S.~Cowperthwaite}
\altaffiliation{NSF GRFP Fellow}
\affiliation{Harvard-Smithsonian Center for
Astrophysics, 60 Garden Street, Cambridge, MA 02138, USA}
\author{M.~Nicholl}
\affiliation{Harvard-Smithsonian Center for
Astrophysics, 60 Garden Street, Cambridge, MA 02138, USA}
\author{V.~A.~Villar}
\altaffiliation{NSF GRFP Fellow}
\affiliation{Harvard-Smithsonian Center for Astrophysics, 60 Garden Street,
Cambridge, MA 02138, USA}
\author{K.~D.~Alexander}
\affiliation{Harvard-Smithsonian Center for
Astrophysics, 60 Garden Street, Cambridge, MA 02138, USA}
\author{P.~K.~Blanchard}
\affiliation{Harvard-Smithsonian Center for
Astrophysics, 60 Garden Street, Cambridge, MA 02138, USA}
\author{T.~Eftekhari}
\affiliation{Harvard-Smithsonian Center for Astrophysics, 60 Garden Street,
Cambridge, MA 02138, USA}
\author{W.~Fong}
\altaffiliation{Hubble Fellow}
\affiliation{Center for Interdisciplinary Exploration and Research in Astrophysics (CIERA) and Department of Physics and Astronomy, Northwestern University, Evanston, IL 60208}
\author{R.~Margutti}
\affiliation{Center for Interdisciplinary Exploration and Research in Astrophysics (CIERA) and Department of Physics and Astronomy, Northwestern University, Evanston, IL 60208}
\author{P.~K.~G.~Williams}
\affiliation{Harvard-Smithsonian Center for
Astrophysics, 60 Garden Street, Cambridge, MA 02138, USA}
\author{J.~Annis}
\affiliation{Fermi National Accelerator Laboratory, P. O. Box 500, Batavia, IL 60510, USA}
\author{D.~Brout}
\affiliation{Department of Physics and Astronomy, University of
Pennsylvania, Philadelphia, PA 19104, USA}
\author{D.~A.~Brown}
\affiliation{Department of Physics, Syracuse University, Syracuse NY 13224, USA}
\author{H.-Y.~Chen}
\affiliation{Department of Astronomy and Astrophysics, University of Chicago, Chicago, Illinois 60637, USA}
\author{M.~R.~Drout}
\altaffiliation{Hubble and Carnegie-Dunlap Fellow}
\affiliation{The Observatories of the Carnegie Institution for Science, 813
Santa Barbara St., Pasadena, CA 91101, USA}
\author{R.~J.~Foley}
\affiliation{Department of Astronomy and Astrophysics, University of California, Santa Cruz, CA 95064, USA}
\author{J.~A.~Frieman}
\affiliation{Fermi National Accelerator Laboratory, P. O. Box 500, Batavia, IL 60510}
\affiliation{Kavli Institute for Cosmological Physics, The University of Chicago, Chicago, IL 60637}
\author{C.~L.~Fryer}
\affiliation{Center for Theoretical Astrophysics, Los Alamos National Laboratory, Los Alamos, NM 87544}
\author{D.~E.~Holz}
\affiliation{Enrico Fermi Institute, Department of Physics, Department of
Astronomy and Astrophysics,\\and Kavli Institute for Cosmological Physics,
University of Chicago, Chicago, IL 60637, USA}
\author{T.~Matheson}
\affiliation{National Optical Astronomy Observatory, 950 North Cherry
Avenue, Tucson, AZ, 85719}
\author{B.~D.~Metzger}
\affiliation{Department of Physics and Columbia Astrophysics Laboratory,
Columbia University, New York, NY 10027, USA}
\author{E.~Quataert}
\affil{Department of Astronomy \& Theoretical Astrophysics Center, University of California, Berkeley, CA 94720-3411, USA}
\author{A.~Rest}
\affiliation{Space Telescope Science Institute, 3700 San Martin Drive, Baltimore, MD 21218, USA}
\affiliation{Department of Physics and Astronomy, The Johns Hopkins University, 3400 North Charles Street, Baltimore, MD 21218, USA}
\author{M.~Sako}
\affiliation{Department of Physics and Astronomy, University of
Pennsylvania, Philadelphia, PA 19104, USA}
\author{D.~M.~Scolnic}
\affiliation{Kavli Institute for Cosmological Physics, The University of Chicago, Chicago,IL 60637, USA}
\author{N.~Smith}
\affiliation{Steward Observatory, University of Arizona, 933 N. Cherry Ave., Tucson, AZ 85721}
\author{M.~Soares-Santos}
\affiliation{Department of Physics, Brandeis University, Waltham, MA 02454,
USA}
\affiliation{Fermi National Accelerator Laboratory, P. O. Box 500, Batavia,
IL 60510, USA}


\begin{abstract}
We present a near-infrared spectral sequence of the electromagnetic counterpart to the binary neutron star merger \gw\ detected by Advanced LIGO/Virgo.  Our dataset comprises 7 epochs of $J+H$ spectra taken with FLAMINGOS-2 on Gemini-South between 1.5--10.5 days after the merger.  In the initial epoch, the spectrum is dominated by a smooth blue continuum due to a high-velocity, lanthanide-poor blue kilonova component.  Starting the following night, all subsequent spectra instead show features that are similar to those 
predicted in model spectra of material with a high concentration of lanthanides, including spectral peaks near 1.07 and 1.55~$\mu$m.  Our fiducial model with 0.04~\msun\ of ejecta, an ejection velocity of $v=0.1c$, and a lanthanide concentration of \xlan=10$^{-2}$ provides a good match to the spectra taken in the first five days, although it over-predicts the late-time fluxes. We also explore models with multiple fitting components, in each case finding that a significant abundance of lanthanide elements is necessary to match the broad spectral peaks that we observe starting at 2.5~d after the merger. These data provide direct evidence that binary neutron star mergers are significant production sites of even the heaviest \rp\ elements.
\end{abstract}


\section{Introduction}

Studies of the solar system abundance patterns more than 60 years ago demonstrated the existence of a large number of heavy isotopes that could not be produced during normal stellar fusion processes and instead required many neutron captures onto heavy nuclei on timescales short relative to the weak decay timescale of a few seconds  \citep{bbfh,cameron57}.  The sites of this phenomenon, given the name \rp\ nucleosynthesis, have remained obscure.
Initial efforts focused on supernovae being the best possible candidates then known for the production of the large neutron fluxes necessary for the \rp\ to occur, but most modern work has found that the conditions in numerical simulations are inadequate to produce the heaviest \rp\ material in the necessary quantities (e.g., \citealt{qw96,thiel11}).  Compact object mergers involving neutron stars were subsequently proposed to be alternative sites that could easily explain the presence of neutron-rich material \citep{ls74,eichler89,frt99}.

\citet{lpminisn} noted that synthesis of \rp\ material would naturally produce a large number of radioactive isotopes whose decay energies could power an optical transient associated with a neutron star merger (see also \citealt{rosswog05}).  This event has become known as a "macronova" \citep{kulkarni05} or "kilonova" \citep{metzger2010}.

Early work assumed that the \rp\ material would have opacities similar to that of iron-peak elements due to a lack of appropriate atomic data (e.g., \citealt{metzger2010}).  This assumption works reasonably well for the light \rp\ nuclei (atomic mass number $A \lesssim 140$).  However, the lanthanide and actinide series of elements, which are produced in the \rp, fundamentally differ from other elements in having their outer valence electrons in the $f$-shell.  This electronic configuration permits a much larger number of transitions within a few eV of the ground state, corresponding to a greatly enhanced opacity to optical photons in material enriched in lanthanides \citep{kasen13}.  Therefore, \rp-enhanced material produced in a kilonova will have its flux pushed into the near-infrared (NIR) relative to the optical peak of material whose opacity is dominated by iron-peak elements \citep{kasen13,barnes13,th13}.  While the treatment of the atomic data remains highly uncertain, there is general agreement that this spectral distinction is a unique signature of material with enhanced \rp\ abundances \citep{kasen17,woll17,tanaka17}.

The detection of the binary neutron star merger \gw\ by Advanced LIGO/Virgo at 12:41 UT (all times in this paper are UT) on 2017 August 17 \citep{ALVgcn,ALVdetection} afforded an extraordinary opportunity to test this scenario.  An optical and infrared counterpart associated with the host galaxy NGC~4993 was independently detected by several teams within the next 12 hours \citep{GWEMcapstone}, including ours \citep{DECAMgcn}.  This optical transient source was variously given the name of SSS17a \citep{SWOPEgcn,SWOPEpaper} and DLT17ck \citep{DLT40gcn,DLT40paper}, as well as an International Astronomical Union name of AT2017gfo.

Immediately following the detection of a transient optical source by our DECam program \citep{DECAMgcn,DECamPaper1}, we triggered our dedicated follow-up spectroscopy program using Gemini-South to search for the direct signatures of \rp\ opacity, the detection of which are reported here.  In Section~\ref{obssec}, we present the observations.  We perform comparisons to a suite of kilonova models in Section~\ref{discsec}, and conclude in Section~\ref{concsec}.
Throughout this work, we assume a distance of 39.5~Mpc and a redshift of $z$=0.009727 for NGC 4993 \citep{rc3,freedman01}, as well as a correction of $E(B-V)=0.105$~mag for Galactic extinction \citep{sf11} using the reddening law of \citet{f99}.

\section{Observations}\label{obssec}
We used FLAMINGOS-2 on the 8~m Gemini-South telescope (F2; \citealt{F2ref}) to obtain a sequence of 7 NIR spectra from 1.5 to 10.5~d after the GW trigger.  All observations used the same setup with the $JH$ grism and $JH$ filter to cover the wavelength range of 0.9--1.8~$\mu$m, at an average dispersion of $\sim$6.5~\AA\ per pixel.  A 4-pixel (0.72$\arcsec$) slit resulted in a mean resolution of $R\approx500$ (600 \kms), although that figure varies significantly across the spectra.  Because of the high airmass of the observations, we were careful to orient the long slit at a position angle of 290$\degr$, which was very close to the parallactic value \citep{alex82}. This was also reasonably close to perpendicular to the offset from the host galaxy, which minimized the amount of galaxy light in the background. The individual exposures were dithered in an ABBA sequence, with typical separations of 30$\arcsec$ between the A and B positions.  As the source was setting at the beginning of each night, the observations usually started in twilight and continued until very high airmass.  A log of observations is given in Table~\ref{spectab}.

\begin{deluxetable}{lDcc}[tb]
\tablecolumns{4}
\tablewidth{0pt}
\tablecaption{Log of F2 Spectroscopic Observations\label{spectab}}
\tablehead{
\colhead{UT Date} &
\multicolumn2c{Epoch}  &
\colhead{Exposure Time} &
\colhead{Airmass} \\
\colhead{(mean)} &
\multicolumn2c{(days)} &
\colhead{(s)} &
\colhead{Range}
}
\decimals
\startdata
2017 Aug 19.03 & 1.5 & $14\times180$ &  $1.8-2.6$ \\
2017 Aug 19.99 & 2.5 & $8\times180$ & $1.4-1.7$ \\
2017 Aug 21.99 & 4.5 & $8\times120$ & $1.5-1.7$ \\
2017 Aug 25.00 & 7.5 & $12\times180$ & $1.6-2.1$ \\
2017 Aug 26.01 & 8.5 & $20\times180$ & $1.7-3.0$ \\
2017 Aug 27.01 & 9.5 & $20\times180$ & $1.7-2.8$ \\
2017 Aug 28.01 & 10.5 & $16\times180$ & $1.9-2.8$ \\
\enddata
\end{deluxetable}

 The two-dimensional frames were processed using standard procedures in the \texttt{gemini} IRAF\footnote{IRAF is distributed by the National Optical Astronomy Observatory, which is operated by the Association of Universities for Research in Astronomy, Inc., under cooperative agreement with the National Science Foundation.} package to perform dark subtraction and apply a flat field correction. A local sky exposure was created from neighboring dithered frames and then scaled and subtracted prior to registration and combination of the images to form the final spectral stack.  
A clear trace from the transient is present at all epochs, even as the contrast with the host galaxy star light became small at later epochs.
The host galaxy contributed a significant amount of emission along the slit, so a local linear background was subtracted during spectral extraction.
The extracted and wavelength-calibrated spectra were flux calibrated and corrected for telluric absorption using observations of A0V telluric standard stars observed at a similar airmass to the object and the \texttt{xtellcor\_general} task in the Spextool IDL package \citep{vacca03,cushing04}.

\begin{figure}
\center
\includegraphics[scale=0.45]{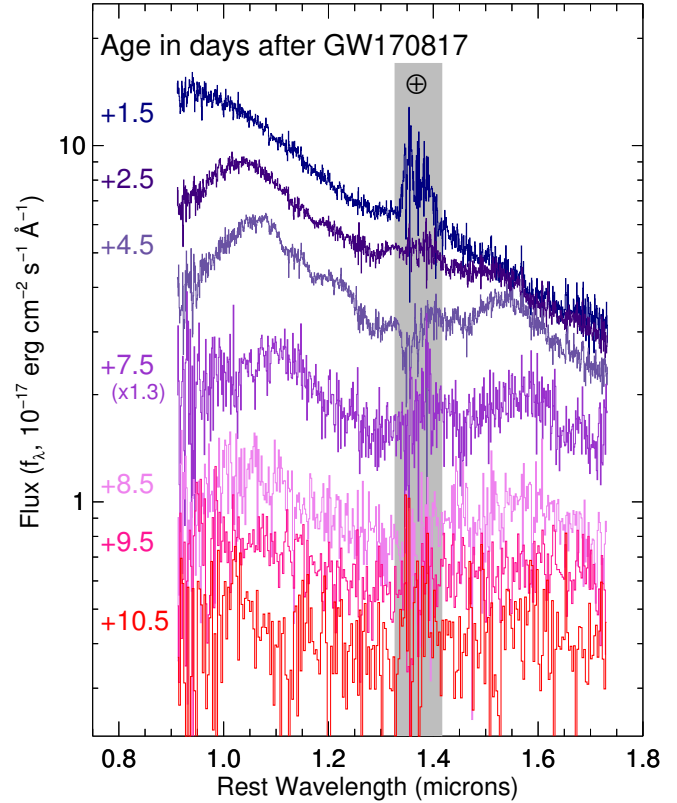}
\caption{NIR spectral sequence of \gw\ from Gemini-South.  Each epoch is labeled with its age in days after the GW trigger.  
The spectra have been de-redshifted, corrected for Galactic extinction, and scaled to match the $H$ photometry \citep{DECamPaper2}.  The $+7.5$~d spectrum 
has been scaled by an additional factor of 1.3 for presentation purposes.  The first three spectra are presented unbinned, but the later ones are binned by increasingly larger factors. The region of strong telluric absorption between $J$ and $H$ bands is indicated by the gray box.}
\label{specfig}
\end{figure}

A pair of $H$ images were taken as part of the acquisition process for each spectral epoch.  The photometry from those images is presented by \citet{DECamPaper2}.  We integrated the spectra over the $H$ bandpass and scaled them to match the photometry.  Our final NIR spectral sequence is presented in Figure~\ref{specfig}.

The initial epoch of NIR spectroscopy, at 1.5~d after the merger, is very smooth. There is a flattening near 1~$\mu$m and a broad shelf present near 1.3--1.4~$\mu$m, which unfortunately coincides with the strong H$_2$O vapor absorption band between $J$ and $H$ bands.  Only one night later, a peak near 1.05~$\mu$m has become prominent and moves redward over the course of the next several nights.  By 4.5~d after the merger, a second peak near 1.55~$\mu$m develops.  This change in appearance, from a smooth blue continuum to a redder continuum with several bumps in addition to the two dominant peaks, is suggestive of a transition in opacity sources of the dominant spectral component over that time.
As the transient fades further, the spectral peaks become less distinct, although that could be a consequence of the decreased signal-to-noise ratio and increased host galaxy contamination over time.

\section{Discussion}\label{discsec}

\subsection{Finding a fiducial model}

To proceed further, we need to consult detailed spectral synthesis models.  We compare to the library of kilonova spectral models produced by \citet{kasen17}, which are an evolved version of previous work \citep{kasen13,barnes13}.  These models assume spherical symmetry, local thermodynamic equilibrium, and uniform abundances, with a density structure that has a central core (density $\propto v^{-1}$) and envelope with a steep density gradient ($\propto v^{-10}$).  The only three tunable parameters are an ejecta mass ($m$), a mean velocity (defined as $v=\sqrt{2K/m}$, where $K$ is the kinetic energy), and a fractional lanthanide abundance (\xlan).  These models were computed for 56 different combinations of these three parameters, with the range of variation motivated by ejecta parameters of various neutron star merger simulations from the literature, such as having ejecta masses around a few hundredths of a solar mass.
We refer the reader to \citet{kasen17} for more details.

We take this library of spectral models, extract the output at the epochs of our NIR spectra, and convert the model luminosities to fluxes at our adopted distance. All model spectra presented in this paper have been smoothed by a 3-pixel boxcar to reduce Monte Carlo sampling noise. All intrinsic spectral features in the data are much broader than this smoothing kernel.  

We make a distinction in subsequent discussion between possible "blue" kilonova components, which are relatively lanthanide poor (\xlan$\lesssim$10$^{-4}$) and "red" kilonova components (\xlan$\gtrsim 10^{-3}$).  In comparison to neutron star merger simulations, the blue component can represent lanthanide-poor material ejected primarily in the polar directions at the interface between the colliding neutron stars or due to neutrino irradiation from the newly-formed hypermassive neutron star (e.g., \citealt{wanajo14,mf14,goriely15}).  The red component is likely to represent material in the equatorial plane, which could originate in tidal tails of the disrupted neutron star, spiral mode instabilities, or winds from an accretion disk (e.g., \citealt{rosswog99,bauswein13,siegel17}).

\begin{figure*}[htb]
\center
\includegraphics[scale=0.75]{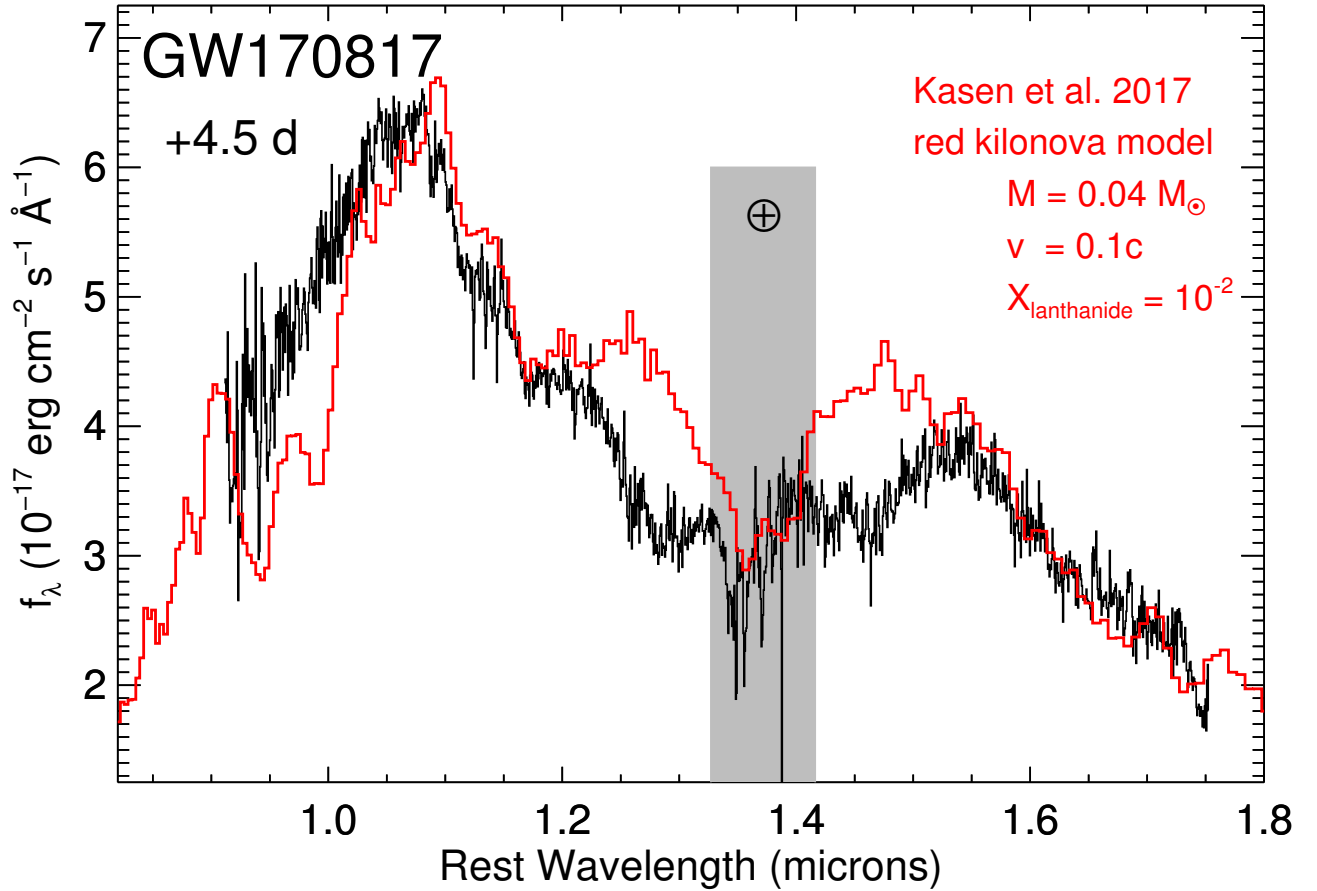}
\caption{The fiducial red kilonova model provides an excellent fit by itself to the day $+4.5$ NIR spectrum, with no adjustments to the flux scale.  The data are in black and the model is in red, with the values of the three main parameters listed on the figure.}
\label{goodfig}
\end{figure*}

We start by examining the 4.5~d spectrum.  We choose this epoch because it has a relatively high signal-to-noise ratio, has clear spectral features present, is sufficiently late that the early emission from the blue kilonova component has faded \citep{DECamPaper2,DECamPaper3}, and yet is well before the ejecta become optically thin and the radiative transfer assumptions might break down.  A simple least-squares statistic finds a best fit among available models with one that has a total mass of 0.04~\msun, expansion velocity 0.1$c$, and a fractional lanthanide abundance \xlan\ of 10$^{-2}$, which is similar to the properties of the red kilonova component independently inferred from the optical and NIR light curves by \citet{DECamPaper2}.  A comparison between the model and data is shown in Figure~\ref{goodfig}.

The agreement between model and data is astonishingly good, especially given the uncertainties in the underlying atomic data, the simplifications inherent to a single-component model, and the lack of any previous NIR spectra for guidance.  We emphasize that we have applied no arbitrary flux scaling factors to the model, and yet it matches the height and wavelength of the 1.07~$\mu$m peak very well.  There is a peak in the model spectrum near 1.5~$\mu$m that is somewhat bluer and higher than the one in the data.  Some of the less prominent features, such as the two shelves of emission between 1.1 and 1.25~$\mu$m on the red side of the 1.07~$\mu$m peak, are also present in both spectra.  The largest disagreement is in the presence of a pair of dips in the model near and blueward of 1.0~$\mu$m that are not visible in the data.  We call this the fiducial model because the excellent agreement between theory and data implies that whatever more complicated scenarios we consider below or can be theoretically proposed, the net effect is to produce emission similar to this single-component model.

The basic features of peaks near 1.1 and 1.55~$\mu$m, with a shelf near 1.25~$\mu$m, are also visible in the favored toy model in the original work of \citet{kasen13}, although those authors also demonstrated that their result depended on the underlying atomic structure model.  The contrast between these features and the expected spectral peak at optical wavelengths for material dominated by iron-peak opacities was regarded as a "smoking gun" of \rp\ nucleosynthesis. The models of \citet{th13} have some roughly similar features, although their spectra are much broader and smoother than in our data and show a much larger drop in flux from 1.0 to 1.6~$\mu$m than the data, which may reflect limitations of the NIR line list used in that work. 

We note that independent sets of kilonova models \citep{woll17,tanaka17}, based on new atomic structure calculations, also reproduce the shift of the observed flux to the NIR when there are high concentrations of lanthanides.  The agreement between various codes about this general trend gives us confidence that this signature of the opacities of \rp\ elements is robust.  
 However, inspection of the figures in those works reveals no clear matches to the spectral sequence as close as the ones we present here. 
  The detailed results of those calculations depend on the assumptions about the masses and compositions of different merger ejecta components. 
 It is unclear at this point whether these detailed spectral differences from different codes represent alternative assumptions about the parameters of the neutron star merger ejecta or differences in the treatment of opacities across the lanthanide series. 

\subsection{Sensitivity to Parameters}

\begin{figure}
\center
\includegraphics[scale=0.45]{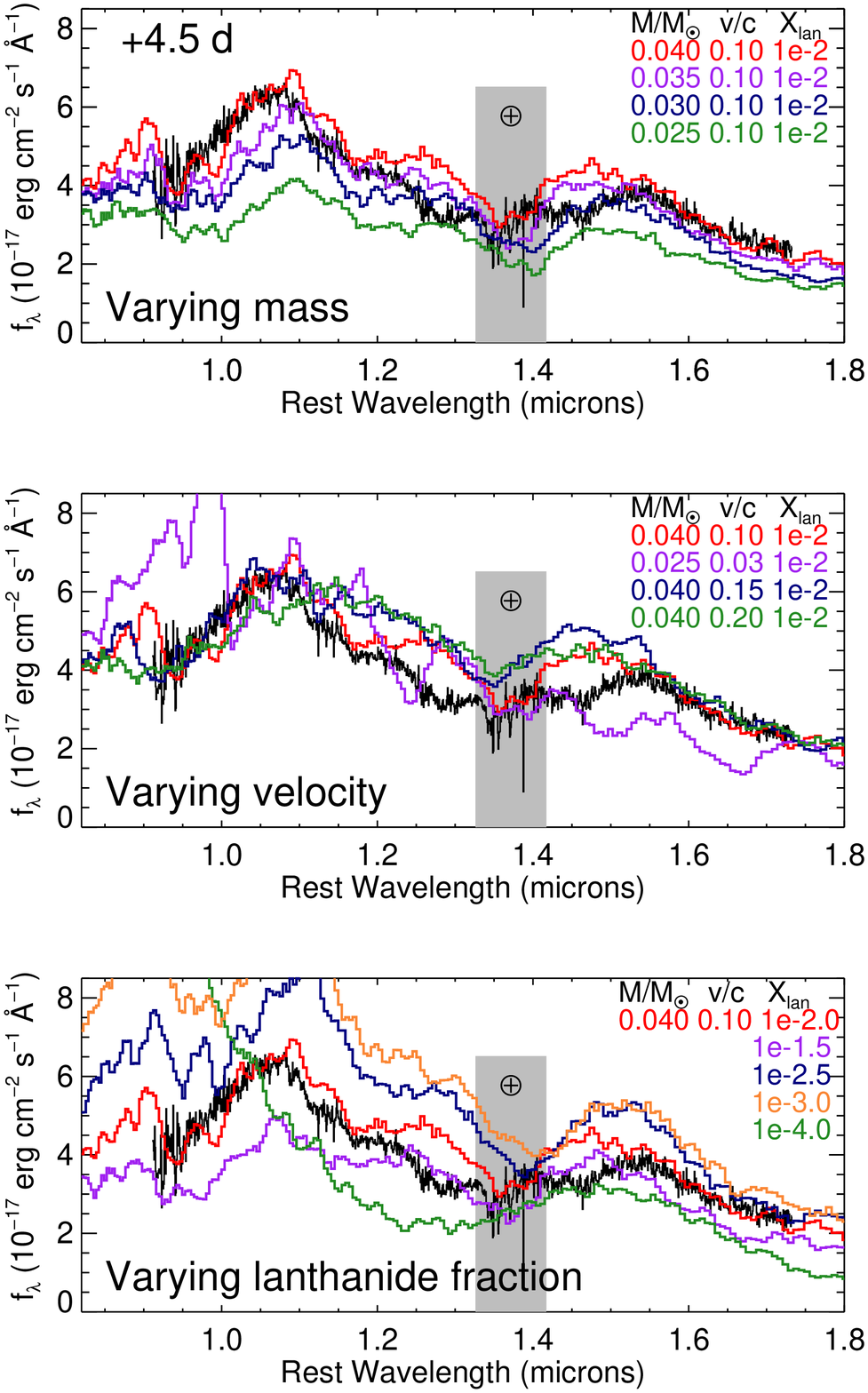}
\caption{The effect of varying model parameters. Top panel: varying the ejecta mass.  Middle panel: the ejecta velocity.  The low-velocity $v=0.03c$ model also has a lower ejecta mass to keep the flux level consistent with the others. Bottom panel: the mass fraction of lanthanide elements. In each panel, the data are shown in black and our fiducial model is red.  The text labels in each panel give the model parameters for the lines of the same color.  }
\label{paramfig}
\end{figure}

We take the excellent agreement between model and data shown in Figure~\ref{goodfig} as a sign that the parameters and the models are at least roughly correct, so now we examine the sensitivity of the model output to the parameter values we have selected.  In the three panels of Figure~\ref{paramfig}, we vary each of the three main parameters in sequence, while holding the other two fixed.
Each of the model spectra also includes a small amount of flux from the same assumed blue kilonova component (not shown, but parameters are discussed below) that contributes a small amount of flux below 1~$\mu$m.

In the top panel, we start by varying the ejecta mass.  As we lower the mass, the overall flux goes down, as expected.  However, the spectra are not simply related by a flux normalization factor. The total mass in the ejecta also affects the diffusion timescale and hence the location of the photosphere within the ejecta.  This, in turn, results in variations in the amount of line blending that shift the wavelengths of the spectral peaks.  Most notably, the 1.07~$\mu$m peak shifts redward at lower ejecta mass.

The ejecta velocity also affects the degree of line blending and smoothness of the spectra.  In the middle panel of Figure~\ref{paramfig}, it is clear that raising the ejecta velocity rounds the tops of the spectral peaks and at $v=0.2c$, the features between 1.1--1.3~$\mu$m are unacceptably washed out relative to the data.  We note that some simulations of the tidal dynamical ejecta find even higher ejecta velocities than this (e.g., \citealt{bauswein13}).  At the other extreme, lowering the ejection velocity results in the major peaks breaking up into a forest of smaller peaks.  The $v=0.03c$ spectrum presented in this panel shows several of these features starting to develop.  Although it is not plotted, by 7.5~d these narrower peaks are predicted to get even more dominant, in contradiction to the smooth broad peaks that we see at that time (Figure~\ref{specfig}).  This is relevant because models invoking strong accretion disk winds (e.g., \citealt{kasen15,siegel17}) predict a range of ejection velocities from 0.03--0.1$c$.  We do not see narrow features expected from material moving as slowly as $v=0.03c$ at any epoch.  If the red kilonova ejecta result from a disk wind, they must be accelerated above this value by, for example, stronger magnetic fields than those previously considered.

Finally, the most important question for the purposes of \rp\ nucleosynthesis is constraining the chemical abundances of the dominant emission component.  In the bottom panel of Figure~\ref{paramfig}, we have adjusted the fractional lanthanide abundance.  If the lanthanide abundance is as low as \xlan=10$^{-4}$, the peak near 1.1~$\mu$m disappears and the model spectra are too blue.  At abundances that are much higher than our fiducial model, the peak near 1.1~$\mu$m is suppressed relative to the one near 1.5~$\mu$m.  Models with \xlan\ between 10$^{-2}$ and 10$^{-3}$ appear to match the overall appearance and ratio of peak heights reasonably well.  The overall flux scaling at lower lanthanide fraction can be somewhat improved by lowering the ejecta mass.

\subsection{Spectral evolution}\label{specevo}

The agreement between data and the model in Figure~\ref{goodfig}, while impressive, is only part of the story.  The next challenge for the models is to self-consistently match the time evolution in the full spectral sequence. 

We find that the fiducial model parameters from Figure~\ref{goodfig} overstate the NIR flux at 7.5~d by almost a factor of 2 and under-predict the flux at 1.5~d.  The latter is not a surprise, as there are independent lines of evidence from the optical photometry \citep{DECamPaper2} and spectra \citep{DECamPaper3} that at the earliest times emission at optical wavelengths is dominated by a lanthanide-poor and high-velocity blue kilonova component.

This observational result, plus theoretical guidance (e.g., \citealt{mf14}), motivates us to consider two-component kilonova models that have both blue and red components with different \rp\ abundances.  We shall proceed by simply summing the flux produced by two separate single-component models.  This is clearly a questionable assumption, but can be justified if either the two kilonova components are spatially disjunct (e.g., a blue polar flow and a red equatorial component) or if one component has a much higher velocity than the other.  We will revisit this below.

\begin{figure*}
\center
\includegraphics[scale=0.45]{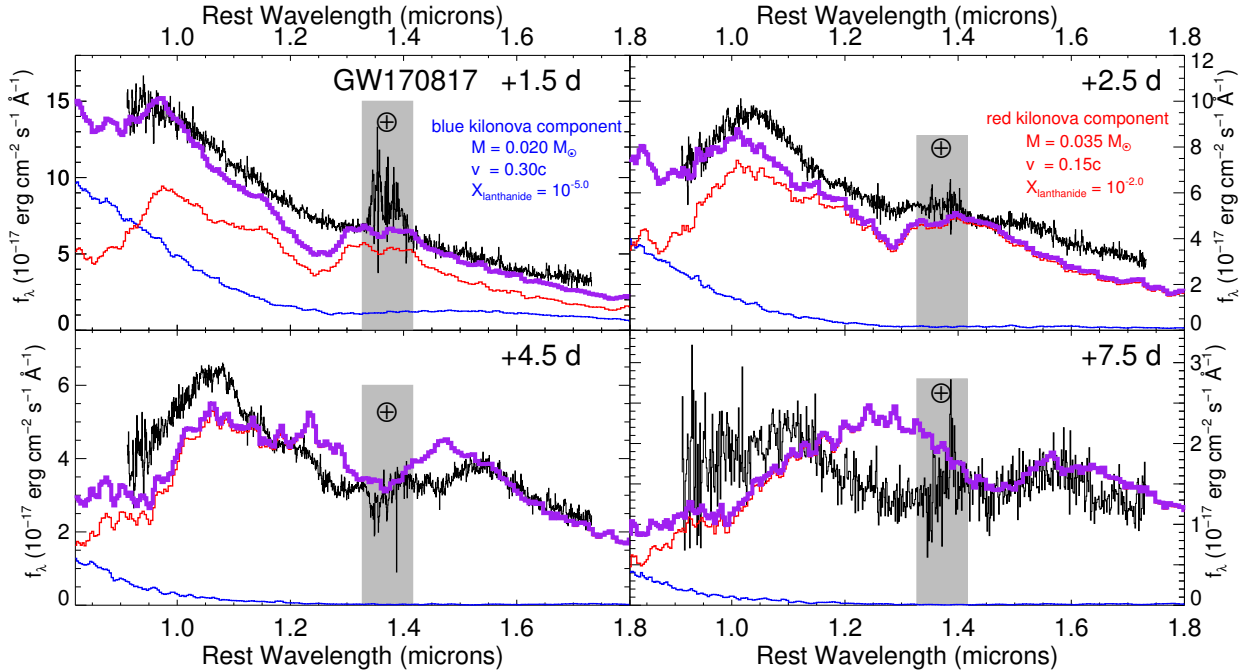}
\caption{A combined blue and red kilonova model that is our best simultaneous fit to the first four epochs of F2 spectra. The data are shown in black. A blue kilonova model is shown in blue and the red kilonova component is in red, with the parameters for each model listed in the upper two panels. The thick purple line represents the sum of the two components.}
\label{evofig}
\end{figure*}

The best simple summation of single-component models relative to the first four spectral epochs that we have found is shown in Figure~\ref{evofig}.  This model has both a blue (\xlan=10$^{-5}$) and a red (\xlan=10$^{-2}$) component, whose sum is shown in purple on each panel of the figure.  While it is not as good a fit to the 4.5~d spectrum as the fiducial model, it alleviates the discrepancies with the flux levels at earlier and later epochs described above.

We start by examining the 1.5~d spectrum because the blue kilonova component only makes a dominant contribution at early times.  The abundances and velocity (0.3$c$) of the blue component are the same found from fitting the optical photometry and spectroscopy \citep{DECamPaper2,DECamPaper3}, and with an ejecta mass (0.02~\msun) between those two estimates.  However, this model still has some inconsistencies.  The actual data are much smoother than the combined model, implying that the photosphere is in high-velocity material with blue colors.  A pure blue kilonova model with a higher ejecta mass (as found by \citealt{DECamPaper3}) would probably be a better fit by itself.  However, any red kilonova component sufficiently luminous to account for the later spectra also contributes some emission at these early times as well.  In addition, this red component imprints a spectral dip near 1.25~$\mu$m on the combined model that is not seen in the data.  One possible resolution of these inconsistencies is if while the photosphere is in the high-velocity blue material at early times, it obscures the contribution from the lower-velocity red kilonova component until the blue kilonova becomes more optically thin at later epochs.

Next, we compare to the later-time spectra (7.5~d).  As described above, the fiducial red kilonova model provides an excellent fit to the 4.5~d spectrum, but over-predicts the late-time flux.  The minimal modifications to the fiducial model to resolve this are to reduce the radiative diffusion timescale by some combination of reducing the ejecta mass (here, 0.035~\msun) and increasing the velocity (0.15$c$).  The flux level at 7.5~d is now acceptable.  However, all model calculations we examined with \xlan=10$^{-2}$ develop a broad third NIR peak between 1.2--1.3~$\mu$m at 7.5~d that is not present in the data. Reducing the lanthanide concentration to \xlan=10$^{-3}$ produces spectra with only two peaks in the NIR at this epoch (at 1.1 and 1.6~$\mu$m), in agreement with the data.
This may be similar to the intermediate or "purple" kilonova component inferred from the photometry by \citet{DECamPaper2}.  However, such a model is too blue at 2.5~d and lacks a spectral peak near 1.05~$\mu$m as strong as that seen in the data.

In summary, we have experimented with resolving the failures of a single-component kilonova model to match the spectral evolution by summing a blue and a red component.  We present such a model in Figure~\ref{evofig} and find that the overall flux level is reasonably good, but that there are several inconsistencies.
We also note that if we were to fit the first four spectra independently, we would find best-fit lanthanide fractions of $\lesssim 10^{-4.5}$, 10$^{-2}$, 10$^{-2}$, and 10$^{-3}$ for days 1.5, 2.5, 4.5, and 7.5, respectively.  No combination of two (or three) components can reproduce this evolution by simple summation, so more complex models are required.  The effects of geometry, self-obscuration, and reprocessing of emission from one component by the other are likely important in reality, but have not been treated in these comparisons. 

\section{Conclusion}\label{concsec}
We have presented a spectral sequence of the NIR counterpart to the binary neutron star merger \gw.  
Our initial spectra are smooth and blue, but develop several broad peaks starting 2.5~d after the merger, indicating a change in dominant opacity sources.
Our highest-quality spectrum is a good match to a fiducial one-component red kilonova model from \citet{kasen17}, despite many theoretical uncertainties in the atomic data and heating rate (e.g., \citealt{barnes16,hoto16}).  We find that in order to reproduce the spectral peaks near 1.05~$\mu$m and 1.55~$\mu$m that dominate the spectra as soon as 2.5~d after the merger, we require kilonova ejecta with a mass of $\sim$0.04~\msun\ and a high lanthanide abundance, \xlan$\approx$10$^{-2}$.  Models that lack a sufficient concentration of material with lanthanide-like opacities result in spectra that are too blue and otherwise inconsistent with the data.  

In order to better match the spectral sequence, we also explored simply adding blue and red kilonova components with different lanthanide abundances.
The properties of the blue component are similar to those found to fit the optical spectra \citep{DECamPaper3}.  In addition, the presence of at least two components with parameters similar to ours was also inferred from fitting the combined optical and NIR light curves \citep{DECamPaper2}. 

  Further progress will require models that have more realistic variations of abundances with velocity within the ejecta. The numbers quoted at the end of Section~\ref{specevo} for the lanthanide abundances in the dominant emission component at each epoch may provide a guide for future work.  
  The excellent agreement between model and data shown in Figure~\ref{goodfig} may be a sign that the properties of the component that dominates the emission at 4.5~d has parameters similar to those of the model.

Matching the derived Galactic \rp\ production rate requires kilonova ejecta masses in the range of 0.01--0.1~\msun\ (e.g., \citealt{FMreview}) if estimates of the binary neutron star merger rate from the Galactic neutron star population are correct \citep{kalogera04,kim15}.  Our best fit two-component model has a total ejecta mass of 0.055~\msun, with 0.035~\msun\ of material enriched in the heavy \rp, squarely in the middle of the range.  Further improvements in the merger rate from Advanced LIGO/Virgo detections and observational constraints on the range of kilonova ejecta masses will greatly reduce the uncertainty in these estimates.  
Future observations using the James Webb Space Telescope will take advantage of the broad spectral range and lack of telluric absorption to enable more complete studies of the merger ejecta components and their detailed abundance patterns.
However, as long as \gw\ is typical of the population, our results demonstrate that binary neutron star mergers do produce sufficient quantities of \rp\ material.

\acknowledgments
We are grateful for the heroic efforts of the entire Gemini-South staff to obtain these observations, which started in twilight each night as the source was setting, particularly M. Andersen, P. Candia, J. Chavez, G. Diaz, R. Diaz, V. Firpo, G. Gimeno, H. Kim, A. Lopez, L. Magill, P. Prado, R. Rutten, R. Salinas, D. Sanmartim, A. Shugart, K. Silva, E. Wenderoth, and the director, L. Ferrarese, for her support and approval of the associated DD program.

Based on observations obtained at the Gemini Observatory 
(Program IDs GS-2017B-Q-8 and GS-2017B-DD-4; PI: Chornock), which is operated by the Association of Universities for Research in Astronomy, Inc., under a cooperative agreement with the NSF on behalf of the Gemini partnership: the National Science Foundation (United States), the National Research Council (Canada), CONICYT (Chile), Ministerio de Ciencia, Tecnolog\'{i}a e Innovaci\'{o}n Productiva (Argentina), and Minist\'{e}rio da Ci\^{e}ncia, Tecnologia e Inova\c{c}\~{a}o (Brazil). 

The Berger Time-Domain Group at Harvard is supported in part by the NSF
through grants AST-1411763 and AST-1714498, and by NASA through grants
NNX15AE50G and NNX16AC22G.

D.K. is supported in part by a Department of Energy Office of Nuclear Physics Early Career Award and by grant DE-SC0017616, and by the Director, Office of Energy Research, Office of High Energy and Nuclear Physics, Divisions of Nuclear Physics, of the U.S. Department of Energy under Contract No.DE-AC02-05CH11231. D.K.'s research was supported in part by the Gordon and Betty Moore Foundation through Grant GBMF5076. This research used resources of the National Energy Research Scientific Computing Center, a DOE Office of Science User Facility supported by the Office of Science of the U.S. Department of Energy under Contract No. DE AC02-05CH11231.

D.A.B. is supported by NSF award PHY-1707954.

R.J.F. thanks the University of Copenhagen, DARK Cosmology Centre, and the Niels Bohr International Academy for hosting him during the discovery of GW170817/SSS17a, where he was participating in the Kavli Summer Program in Astrophysics, "Astrophysics with gravitational wave detections."  This program was supported by the Kavli Foundation, Danish National Research Foundation, the Niels Bohr International Academy, and the DARK Cosmology Centre.
The UCSC group is supported in part by NSF grant AST--1518052, the Gordon \& Betty Moore Foundation, the Heising-Simons Foundation, generous donations from many individuals through a UCSC Giving Day grant, and from fellowships from the Alfred P.\ Sloan Foundation and the David and Lucile Packard Foundation to R.J.F.

M.R.D. is supported by
NASA through Hubble Fellowships awarded by the Space Telescope Science
Institute, which is operated by the Association of Universities for
Research in Astronomy, Inc., for NASA, under contract NAS 5-26555.

\facilities{Gemini:South (FLAMINGOS-2)} 
\software{IRAF, IDL, Spextool}

\end{document}